\RequirePackage{fix-cm}
\documentclass[smallextended]{svjour3}       
\smartqed  
\usepackage{graphicx}
\begin{document}

\title{Correlation quantum beats induced by non-Markovian effect}

\author{ }

\institute{  }

\date{Received: date / Accepted: date}

\maketitle

\begin{abstract}
For two qubits independently coupled to their respective structured
reservoirs (Lorentzian spectrum), quantum beats for entanglement and
discord are found which are the result of quantum interference
between correlation oscillations induced by local non-Markovian
environments. We also discuss the preservation of quantum
correlations by the effective suppression of the spontaneous
emission.
\keywords{Quantum beat \and Entanglement \and Quantum discord \and Non-Markovian effect}
\end{abstract}

\section{Introduction}
\label{intro}
Quantum entanglement and discord, two different measures of quantum
correlation, have been receiving much attention recent years.
Interest that people pour on entanglement lies in mainly two
aspects. One is the important role of entanglement in quantum
computation and quantum information processing \cite{Nielsen}, such
as quantum teleportation, quantum dense coding, quantum
cryptography, remote quantum-state preparation \cite{Berry}, quantum
remote control \cite{Huelga1}, and distributed quantum computation
\cite{Cirac}, etc. The other lies in the understanding of nature of
non-locality in quantum mechanics \cite{Einstein}. Though quantum
entanglement not necessarily leads to non-locality, quantum
non-locality does indeed arise from measurement correlations on
quantum entangled states. Quantum discord, a relatively new concept
born at the turn of this century, is regarded as the measure of all
nonclassical correlations in a bipartite system \cite{Ollivier},
being the entanglement a particular case of it. As shown in
\cite{Datta,Lanyon}, deterministic quantum computation with one pure
qubit may be carried out in the situation of no entanglement but
discord. In addition, quantum discord is also used in studies of
quantum phase transition \cite{Dillenschneider,Sarandy}, estimation
of quantum correlations in the Grover search algorithm \cite{Cui},
and to characterize the class of initial system-bath states for
which the quantum dynamics is equivalent to a completely positive
map \cite{Shabani}.

Realistic quantum systems cannot avoid interactions with their
environments, which can alter the properties of quantum correlation.
In the last decade, the influences of Markovian environments on
quantum entanglement have been investigated extensively \cite{Ann}.
An interesting phenomenon, named "entanglement sudden death" (ESD)
\cite{Yu} for a pair of entangled qubits exposed to local Markovian
environments, was found, which triggered reams of related works.
Different from entanglement, quantum discord in similar conditions
decays only in asymptotic time \cite{Werlang}, which signifies that
quantum discord is more robust against Markovian noise than
entanglement.

More recently, much attention and interest have been devoted to the
study of non-Markovian dynamics of open systems
\cite{Breuer,Rivas,Wolf,Dijkstra,Jing,Haikka,Xu1}. As people have
found that many relevant open systems, such as quantum optical
system \cite{Breuer3}, quantum dot \cite{Kubota}, color-core spin in
semiconductor \cite{Kane}, as well as quantum chemistry \cite{Shao}
and biological systems \cite{Chin}, require to be treated in
non-Markovian dynamics. The properties of entanglement evolution for
a pair of entangled qubits exposed to local non-Markovian
environments have been studied already
\cite{Bellomo,Bellomo1,Tong,Xiao}. Entanglement trapping in
engineering structured environment has been proposed. In this paper,
we further investigate the time-evolution properties of quantum
correlation (including entanglement and discord) in this physical
system. We find that a single non-Markovian environment only induces
correlation oscillation, but the cooperation of two local
non-Markovian environments can lead to quantum interference between
correlation oscillations. As a result, an interesting physical
phenomenon--correlation quantum beat--is observed. We will also
highlight the mechanism for the preservation of quantum correlations
in the considered system.

The paper is organized as follows. In Sec. 2, we present the exact
dynamics for a two-qubit system coupled to local non-Markovian
environments. And in Secs. 3 and 4, we study the quantum beats
for entanglement and discord respectively for the studied open
system. Sec. 5 is devoted to the problem of preservation of quantum
correlations for the considered system. A conclusion is arranged in
Sec. 6.

\section{Non-Markovian dynamics}
\label{sec:1}
In this part, we present the exact dynamics of the open quantum
system considered in this paper. We consider two noninteracting
qubits \emph{A} and \emph{B}, each of them locally coupled to a
vacuum reservoir via Jaynes-Cummings model,
\begin{equation}
H=\omega_{0}\sigma_{+}\sigma_{-}+\sum_{k}\omega_{k}b^{+}_{k}b_{k}+\sum_{k}(g_{k}\sigma_{+}b_{k}+h.c.).
\end{equation}
Here $\omega_{0}$ is the transition frequency of qubit and
$\sigma_{\pm}$ are the corresponding raising and lowering operators,
while the index $k$ labels the field modes of the reservoir with
frequencies $\omega_{k}$, creation and annihilation operators
$b^{+}_{k}$, $b_{k}$ and coupling constants $g_{k}$. Eq. (1) is one
of the most fundamental system-environment interaction models which
is extensively used in modeling dynamics of open quantum systems
\cite{Haikka,Breuer3,Bellomo,Laine}. According to the method
presented in \cite{Bellomo}, the dynamics of the two-qubit system
can be obtained simply from the one of the individual qubits. The
exact dynamics of each qubit can be described as \cite{Laine},
\begin{equation}
\begin{array}{l}
  \rho_{11}(t)=|G(t)|^{2}\rho_{11}(0), \\
  \rho_{00}(t)=\rho_{00}(0)+[1-|G(t)|^{2}]\rho_{11}(0), \\
  \rho_{10}(t)=G(t)\rho_{10}(0),\\
  \rho_{01}(t)=G^{*}(t)\rho_{01}(0).
\end{array}
\end{equation}
Where $\rho_{ij}=\langle i|\rho |j\rangle$ with $|0\rangle$ and
$|1\rangle$ being ground and excited states of qubit. The function
$G(t)$ is defined as the solution of the integro-differential
equation,
\begin{equation}
\frac{d}{dt}G(t)=-\int_{0}^{t}dt_{1}f(t-t_{1})G(t_{1}),
\end{equation}
with initial condition $G(0)=1$. The kernel $f(t-t_{1})$ denotes the
two-point reservoir correlation function which is the Fourier
transformation of the spectral density$J(\omega)$,
\begin{equation}
f(t-t_{1})=\int d\omega
J(\omega)\exp[i(\omega_{0}-\omega)(t-t_{1})],
\end{equation}
with $\omega_{0}$ being the transition frequency of qubit.

Although the process applies to any form of initial correlation
state of two-qubit system, for simplicity, we only consider two
types of initial Bell-like states,
\begin{equation}
|\psi\rangle=a|00\rangle+b e^{i\theta}|11\rangle
\end{equation}
and
\begin{equation}
|\phi\rangle=\alpha|01\rangle+\beta e^{i\delta}|10\rangle.
\end{equation}
Where $a$, $b$, $\alpha$, $\beta$, $\theta$, $\delta$ are real and
normalized as $a^{2}+b^{2}=1$ and $\alpha^{2}+\beta^{2}=1$. Owing to
the influence of the environment, the coherence of the system will
vanish asymptotically. Thus, the state of the system at any time $t$
would be a mixed state. Under the product basis $\{|00\rangle,
|01\rangle, |10\rangle, |11\rangle\}\equiv \{|1\rangle, |2\rangle,
|3\rangle, |4\rangle\}$, the dynamical state of the two-qubit system
can be written in the form $\rho(t)=[\rho_{ij}(t)]$ with
$i,j=1,2,3,4$. By the method presented in \cite{Bellomo}, combined
with the dynamical evolution of eq.(2) for single open qubit, we
find that the density matrix of the open two-qubit system only has
the following nonzero elements,
\begin{equation}
\begin{array}{l}
\rho_{11}(t)=a^{2}+(1-|G_{A}|^{2})(1-|G_{B}|^{2})b^{2},\\
\rho_{22}(t)=(1-|G_{A}|^{2})|G_{B}|^{2}b^{2},\\
\rho_{33}(t)=|G_{A}|^{2}(1-|G_{B}|^{2})b^{2},\\
\rho_{44}(t)=|G_{A}|^{2}|G_{B}|^{2}b^{2},\\
\rho_{14}(t)=G_{A}^{*}G_{B}^{*}abe^{-i\theta},\\
\rho_{41}(t)=G_{A}G_{B}abe^{i\theta},
\end{array}
\end{equation}
for initial state $|\psi\rangle$, and
\begin{equation}
\begin{array}{l}
\rho_{11}(t)=(1-|G_{B}|^{2})\alpha^{2}+(1-|G_{A}|^{2})\beta^{2},\\
\rho_{22}(t)=\alpha^{2}|G_{B}|^{2}, \\
\rho_{33}(t)=\beta^{2}|G_{A}|^{2},\\
\rho_{23}(t)= \alpha\beta e^{-i\delta}G_{A}^{*}G_{B},\\
\rho_{32}(t)= \alpha\beta e^{i\delta}G_{A}G_{B}^{*},
\end{array}
\end{equation}
for initial state $|\phi\rangle$. All other matrix elements are
zero. Here we use $G_{A}\equiv G_{A}(t)$ and $G_{B}\equiv G_{B}(t)$
to denote the $G(t)$ function for qubits $A$ and $B$ respectively,
which are determined by eq.(3). In the following sections, we will
study the dynamical properties of quantum correlations based on
these equations.

\section{Entanglement quantum beat}
\label{sec:2}
Using the results presented in above section, one can study the
evolution of quantum entanglement for given initial states. We use
Wootters concurrence \cite{Wootters} as the measure of quantum
entanglement. For a two-qubit system with density matrix $\rho$, the
concurrence $C$ is defined as
\begin{equation}
C=\max\{0,\lambda_{1}-\lambda_{2}-\lambda_{3}-\lambda_{4}\}.
\end{equation}
Here $\lambda_{1}\geq\lambda_{2}\geq\lambda_{3}\geq\lambda_{4}$ are
the square roots of the eigenvalues of the matrix $R=\rho
\widetilde{\rho}$, with
$\widetilde{\rho}=\sigma_{y}^{(1)}\otimes\sigma_{y}^{(2)}\rho^{*}\sigma_{y}^{(1)}\otimes\sigma_{y}^{(2)}$
and the sign ``$\ast$" standing for complex conjugate. For the
states given by eqs.(7) and (8), the corresponding concurrences are
\begin{equation}
C_{\psi}(t)=\max\{0,2|G_{A}G_{B}ab|-2b^{2}|G_{A}G_{B}|\sqrt{(1-|G_{A}|^{2})(1-|G_{B}|^{2})}\}
\end{equation}
and
\begin{equation}
C_{\phi}(t)=2|\alpha\beta G_{A}G_{B}|,
\end{equation}
respectively. The notable characteristic is that the dynamical
entanglement is independent of the phase angles $\theta$, $\delta$
in the initial Bell-like states. In addition, the evolution of state
$|\psi\rangle$ can happen entanglement sudden death (ESD) when the
condition $|a|-|b|\sqrt{(1-|G_{A}|^{2})(1-|G_{B}|^{2})}<0$ is
fulfilled.

The results of eqs. (10) and (11) apply to, in general, any
structured environment. Here we assume that two identical qubits $A$
and $B$ (having equal transition frequency $\omega_{0}$  and free
decay rate $\gamma_{0}$) are respectively plug into different
imperfect optical cavities, so that the spectrum may be
characterized by the Lorentzian distribution,
\begin{equation}
J_{j}(\omega)=\frac{{{\gamma}_{0}}{\lambda}^2}{2{\pi}[({\omega}_{0}-\omega-\Delta_{j})^2+{\lambda}^2]}.
\end{equation}
Where $\Delta_{j}=\omega_{0}-\omega_{cj}$ with $j=A,B$ is the
frequency detuning between qubit $j$ and the corresponding
cavity-mode $\omega_{cj}$. We assume that the two cavities have
equal photon-leakage rate $\lambda$. For this Lorentzian spectrum,
the correlation function in eq.(3) can be easily evaluated as
\begin{equation}
f_{j}(t-t_1)=\frac{1}{2}{\gamma}_0{\lambda}e^{-(\lambda-i\Delta_{j})(t-t_1)},
\end{equation}
and thus the function $G(t)$ is
\begin{equation}
G_{j}(t)=e^{-(\lambda-i\Delta_{j})t/2}[\cosh(\frac{d_{j}t}{2})+\frac{\lambda-i{\Delta_{j}}}{d}\sinh(\frac{d_{j}t}{2})],
\end{equation}
with
$d_{j}=\sqrt{(\lambda-i{\Delta_{j}})^2-2{{\gamma}_0}{\lambda}}$.
Inserting this expression into eqs.(10) and (11), we can obtain the
dynamical entanglement for the given initial states $|\psi\rangle$
and $|\phi\rangle$.

For the Lorentzian spectrum, two typical coupling regimes--weak and
strong regimes--can be distinguished customarily. For the weak
coupling regime, $\gamma_{0}<\lambda/2$, the correlation time
$\tau_{B}\simeq \lambda^{-1}$ of reservoir is smaller than the
relaxation time $\tau_{R}\simeq \gamma_{0}^{-1}$ of the qubit. In
this regime, the qubit dynamics behaves like essentially
Markovianity and the quantum entanglement between qubits $A$ and $B$
reduces monotonously. In contrast, for the strong coupling regime,
$\gamma_{0}>\lambda/2$, the correlation time of reservoir is greater
than or of the same order as the relaxation time of the qubit. In
this case, the non-Markovian effect may cause the quantum
entanglement between the two qubits to rise, or even to form
entanglement oscillations. In particular, we find that when the
detunings $\Delta_{A}$ and $\Delta_{B}$ are both large and have tiny
difference, then the time evolution for the concurrences given by
eqs.(10) and (11) will happen quantum beat, an interesting and
fundamental quantum phenomenon encountered in quantum optics. We
show this in Fig.1, where the initial Bell states are chosen, but
for other non-maximally entangled states, similar phenomenon of
quantum beats may also exist. It is worthwhile to point out that
these quantum beats originate purely from non-Markovian effect, as
no any direct or mediated interaction exists between the two qubits.
This is different from the case \cite{Francica} of two qubits
coupled to a common cavity, where the two qubits can interact
through common reservoir. In addition, Fig.1 also shows that the
entanglement decay for state $|\psi\rangle$ is faster than that for
state $|\phi\rangle$, which indicates that the later is more robust
when exposing to the considered reservoirs. The physical reason is
visible: In state $|\phi\rangle$, at most one of the qubits is in
excited state so that the rate of spontaneous emission is smaller.
While in state $|\psi\rangle$, two qubits can be simultaneously in
excited state, leading to stronger spontaneous emission.

\begin{figure}
  \includegraphics[width=0.75\textwidth]{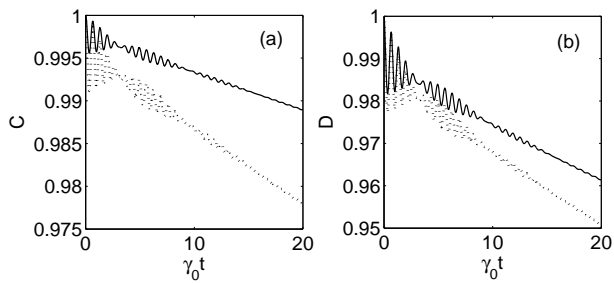}
\caption{Time evolution of concurrence for initial states $|\phi\rangle=1/\sqrt{2}[|01\rangle+e^{i\delta}|10\rangle]$ (solid line),
  and $|\psi\rangle=1/\sqrt{2}[|00\rangle+e^{i\theta}|11\rangle]$ (dot line). Where $\lambda=0.2\gamma_{0}$, $\Delta_{A}=50\lambda$, $\Delta_{B}=45\lambda$.}
\label{fig:1}
\end{figure}

Physically, the entanglement quantum beat is the result of quantum
interference of the dynamical entanglements. If only one of the
qubits couples to environment and the other is immune from any
noise, then the non-Markovian effect only leads the dynamical
entanglement between the two qubits to oscillate but not to happen
quantum beat. This features are shown in Fig.2 (top row), where the
solid line denotes the evolution of concurrence when only qubit $A$
couples to environment with detuning $\Delta_{A}=50\lambda$, and the
dot line for the case where only qubit $B$ couples to environment
with detuning $\Delta_{B}=45\lambda$. The different detunings
$\Delta_{A}$ and $\Delta_{B}$ leads to the tiny difference in
frequencies for the two corresponding entanglement oscillations. In
this way, when the two qubits couple simultaneously to their local
environments, the entanglement oscillations produce interference,
leading to entanglement quantum beat. The phenomenon is similar to
the coherence superposition of two Mechanical oscillations with tiny
frequency difference to form classical beat, except that the total
dynamical entanglement is not simply the sum of the two components
of the dynamical entanglements depicted in Fig.2.

In the above discussion, we produce the tiny frequency difference
for the two entanglement oscillations by different cavity
frequencies $\omega_{cj}$. In fact, we may also realize the
entanglement quantum beat through other parameter control, for
example by adjusting the transition frequencies of the qubits or the
Lorentzian widths $\lambda_{j}$ of the cavities. In this sense, the
quantum beat may be used to indicate the differences of the
cavity-mode frequencies, the qubit-transition frequencies or the
cavity-decay rates, according to different situations.

\begin{figure}
  \includegraphics[width=0.75\textwidth]{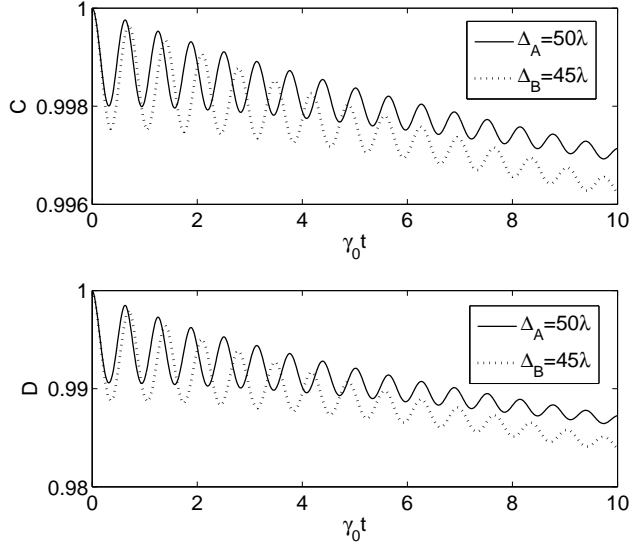}
\caption{Time evolution of concurrence (top row) and discord (bottom row) for initial state
  $|\phi\rangle=1/\sqrt{2}[|01\rangle+e^{i\delta}|10\rangle]$.
  Where $\lambda=0.2\gamma_{0}$.}
\label{fig:2}
\end{figure}

\section{Quantum beat for discord}
\label{sec:3}

Besides the entanglement quantum beat, we find that the time
evolution of quantum discord also appears similar phenomenon in
considered system. In order to demonstrate this, let us first
briefly review the notion of quantum discord. For a bipartite
quantum system, the quantum mutual information between the two
subsystems $A$ and $B$ is given by
\begin{equation}
\mathcal{I}(\rho_{AB})=S(\rho_{A})+S(\rho_{B})-S(\rho_{AB}),
\end{equation}
where $S(\rho)=-{\tt Tr}(\rho\log\rho)$ is the von Neumann entropy
of the density matrix $\rho$. The quantum mutual information has
fundamental physical significance, and is usually used as a measure
of total correlations that include quantum and classical ones. The
classical correlation may be defined in terms of projective
measurement. Suppose we perform a set of projective measurement
$\{\Pi^{B}_{k}\}$ on the subsystem $B$, then the probability for
measurement outcome $k$ may be reads $p_{k}={\tt
Tr}_{AB}[(I^{A}\otimes\Pi^{B}_{k})\rho_{AB}(I^{A}\otimes\Pi^{B}_{k})]$
with $I^{A}$ the identity operator for subsystem $A$. After this
measurement, the state of subsystem $A$ is described by the
conditional density operator $\rho_{A|k}={\tt
Tr}_{B}[(I^{A}\otimes\Pi^{B}_{k})\rho_{AB}(I^{A}\otimes\Pi^{B}_{k})]/p_{k}$.
We define the upper limit of the difference between the von Neumann
entropy $S(\rho_{A})$ and the based-on-measurement quantum
conditional entropy $\sum_{k}p_{k}S(\rho_{A|k})$ of subsystem $A$,
i.e.,
\begin{equation}
\mathcal{Q}(\rho_{AB})=\sup_{\{\Pi^{B}_{k}\}}\left[S(\rho_{A})-\sum_{k}p_{k}S(
\rho_{A|k})\right],
\end{equation}
as the classical correlation of the two subsystems. The maximum is
taken for all possible projective measurements. In the end, the
quantum discord is defined as the difference between the total and
classical correlations \cite{Ollivier},
\begin{equation}
D(\rho_{AB})=\mathcal{I}(\rho_{AB})-\mathcal{Q}(\rho_{AB}).
\label{defiofQD}
\end{equation}

Though in general the calculation of quantum discord is formidable,
for X states of two-qubit system, an analytical method has been
presented \cite{Mazhar}. Employing this method, a straightforward
calculation gives the quantum discord of the state of eqs.(7) as,
\begin{equation}
D_{\psi}(\rho)=S(\rho^{B})+\sum_{j=0}^{3}\lambda_{j}\log_{2}\lambda_{j}+\min\{S_{1},S_{2}\}.
\end{equation}
Where
$S(\rho^{B})=-[(\rho_{11}+\rho_{33})\log_{2}(\rho_{11}+\rho_{33})+(\rho_{22}+\rho_{44})\log_{2}(\rho_{22}+\rho_{44})]$
is the Von Neumann entropy of qubit $B$, given the two-qubit state
described by eq.(7). And
$\lambda_{0}=\frac{1}{2}[(\rho_{11}+\rho_{44})+\sqrt{(\rho_{11}-\rho_{44})^{2}+4|\rho_{14}|^{2}}]$,
$\lambda_{1}=\frac{1}{2}[(\rho_{11}+\rho_{44})-\sqrt{(\rho_{11}-\rho_{44})^{2}+4|\rho_{14}|^{2}}]$,
$\lambda_{2}=\rho_{22}$, $\lambda_{3}=\rho_{33}$, are the
eigenvalues of density matrix given by eq.(7).
$S_{1}=(\rho_{22}+\rho_{44})S(\rho_{0})+
(\rho_{11}+\rho_{33})S(\rho_{1})$ with
$S(\rho_{0})=-\frac{1-\eta}{2}\log_{2}\frac{1-\eta}{2}-\frac{1+\eta}{2}\log_{2}\frac{1+\eta}{2}$,
$S(\rho_{1})=-\frac{1-\eta'}{2}\log_{2}\frac{1-\eta'}{2}-\frac{1+\eta'}{2}\log_{2}\frac{1+\eta'}{2}$,
$\eta=|\rho_{22}-\rho_{44}|/(\rho_{22}+\rho_{44})$,
$\eta'=|\rho_{11}-\rho_{33}|/(\rho_{11}+\rho_{33})$. And
$S_{2}=-\frac{1-\epsilon}{2}\log_{2}\frac{1-\epsilon}{2}-\frac{1+\epsilon}{2}\log_{2}\frac{1+\epsilon}{2}$
with
$\epsilon=\sqrt{(\rho_{11}+\rho_{22}-\rho_{33}-\rho_{44})^{2}+4|\rho_{14}|^{2}}$.

Similarly, the quantum discord for the state of eq.(8) can be
written as,
\begin{equation}
D_{\phi}(\rho)=S(\rho^{B})+\rho_{11}\log_{2}\rho_{11}+(\rho_{22}+\rho_{33})\log_{2}(\rho_{22}+\rho_{33})+\min\{S_{1},S_{2}\}.
\end{equation}
Here
$S(\rho^{B})=-[(\rho_{11}+\rho_{33})\log_{2}(\rho_{11}+\rho_{33})+\rho_{22}\log_{2}\rho_{22}]$
is Von Neumann entropy of qubit $B$, given the two-qubit state
described by eq.(8).
$S_{1}=(\rho_{11}+\rho_{33})[-\frac{1-\Lambda}{2}\log_{2}\frac{1-\Lambda}{2}-\frac{1+\Lambda}{2}\log_{2}\frac{1+\Lambda}{2}]$
with $\Lambda=|\rho_{11}-\rho_{33}|/(\rho_{11}+\rho_{33})$. And
$S_{2}=-\frac{1-\Lambda'}{2}\log_{2}\frac{1-\Lambda'}{2}-\frac{1+\Lambda'}{2}\log_{2}\frac{1+\Lambda'}{2}$
with
$\Lambda'=\sqrt{(\rho_{11}+\rho_{22}-\rho_{33})^{2}+4|\rho_{23}|^{2}}$.

In Fig.3, we show the time evolution of quantum discord according to
eqs. (18) and (19). On the whole, this figure is similar to Fig.1.
The discords for the given states decrease oscillatorily and the
behavior of quantum beat is visible. This quantum beat is also
purely from non-Markovian effect, because quantum discord for
Markovian processes only reduces monotonously \cite{Werlang}. The
physical mechanism is also due to the quantum interference of
discords: When only one of the qubits couples to environment, the
quantum discords fall oscillatorily and no quantum beat occurs [ see
Fig.2 (bottom row)]; But when the two qubits couple simultaneously
to their local environments, the quantum interference lead to
discord quantum beat. Also, the discord for state $|\psi\rangle$
decays faster than that for state $|\phi\rangle$, duo to the
difference of decay rates of the two states. The visible difference
between figs. 3 and 1 is that, in the same parameter conditions,
discord decays more rapidly.

\begin{figure}
  \includegraphics[width=0.75\textwidth]{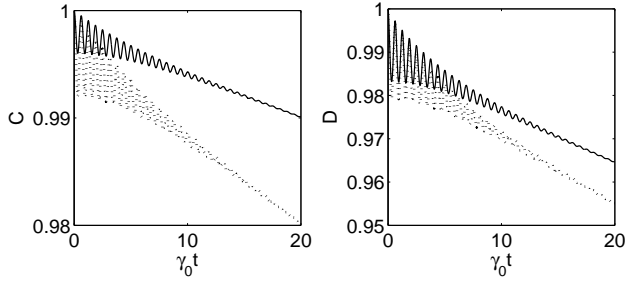}
\caption{Time evolution of quantum discord for initial states $|\phi\rangle=1/\sqrt{2}[|01\rangle+e^{i\delta}|10\rangle]$ (solid line),
  and $|\psi\rangle=1/\sqrt{2}[|00\rangle+e^{i\theta}|11\rangle]$ (dot line). Where $\lambda=0.2\gamma_{0}$, $\Delta_{A}=50\lambda$, $\Delta_{B}=45\lambda$.}
\label{fig:3}
\end{figure}

\section{Protection of quantum correlation}
\label{sec:4}

The protection of
quantum correlation is of great importance to quantum information
processing. Most works \cite{Bellomo,Bellomo1,Tong,Xiao} have been
done for the protection of entanglement in non-Markovian
environments. Here, we mainly discuss the problem of preserving
discord. It is well known that the radiative properties of an atom
in a cavity will change fundamentally compared to in free space.
Spontaneous emission may be decreased \cite{Kleppner} when the atom
is set detuning enough from the cavity. Especially, if the
transition frequency of the atom lies below the fundamental
frequency of the cavity, spontaneous emission is significantly
inhibited. Conversely, if the atom is set to be resonant with the
cavity, then the spontaneous emission will be quicken vastly
\cite{Goy,Purcell}.

The enhancement and suppression of spontaneous emission of atoms in
cavities may be used to effectively manipulate quantum correlation,
even in the non-Markovian environments. For example, to protect
quantum correlation between two qubits, one can put the two qubits
into two separate vacuum cavities, as implemented in the previous
sections. By adjusting the cavity frequencies well mistuned from the
qubit transitions, the spontaneous emission of the qubits is then
inhibited and the quantum correlation between them is protected
effectively. In Fig. 4, we depict the time evolution of discord
between the two qubits, according to eqs.(18) and (19). [The
evolution of concurrence is similar but decreases slower.] For
simplicity, we set the two qubit to have equal detuning,
$\Delta_{A}=\Delta_{B}\equiv \Delta$ [For different detunings, the
quantum beat may occur as seen in previous sections]. One may see
clearly from this figure that in the resonant regime, the discord
decreases rapidly. However, for large detunings, the discord can be
protected much effectively.

\begin{figure}
  \includegraphics[width=0.75\textwidth]{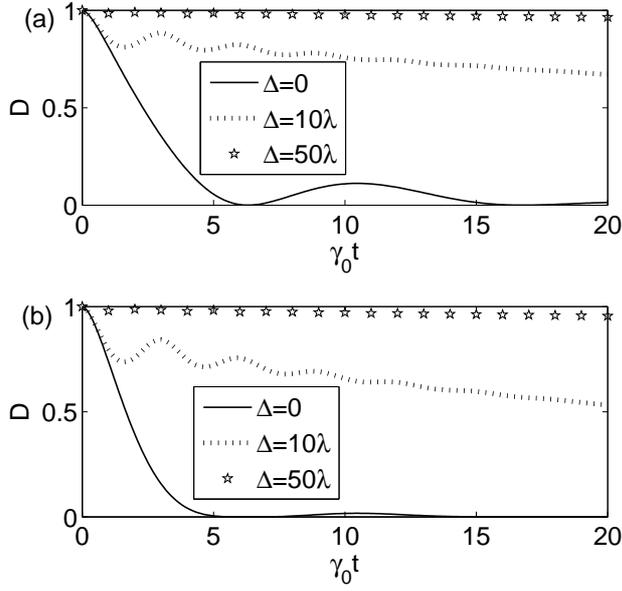}
\caption{Time evolution of quantum discord for different detunings $\Delta$ and for initial states: (a) $|\phi\rangle=1/\sqrt{2}[|01\rangle+e^{i\delta}|10\rangle]$,
  and (b) $|\psi\rangle=1/\sqrt{2}[|00\rangle+e^{i\theta}|11\rangle]$. Where $\lambda=0.2\gamma_{0}$.}
\label{fig:4}
\end{figure}

Two significant elements for the protection of quantum correlations
in the above setup deserve to mention. One is the amount of
frequency detuning between qubit and cavity mode. The larger the
detuning is, the smaller the rate of spontaneous emission is, and
the slower the quantum correlation reduces. This feature has been
observed already from Fig.4. Another significant element is the
monochromaticity of the cavity mode. For given qubit-cavity
detuning, the well the monochromaticity of the cavity mode is, the
smaller the rate of spontaneous emission is, and the slower the
quantum correlation decreases. This feature is shown in Fig.5. [Note
that the two features mentioned here are also true for concurrence.]
Especially, in the limit that the cavity mode is completely
monochromatic [which mean the Lorentzian width in eq.(12)
$\lambda\rightarrow0$ ] and for non-resonant case, one has
$d_{j}\rightarrow i|\Delta|$ and $|G_{j}|\rightarrow1$. Then eqs.(7)
and (8) return respectively to their initial states of eqs. (5) and
(6). This implies that qubits in non-leakage and non-resonant
cavities, the spontaneous emissions are completely suppressed and
their quantum states are preserved exactly.

\begin{figure}
  \includegraphics[width=0.75\textwidth]{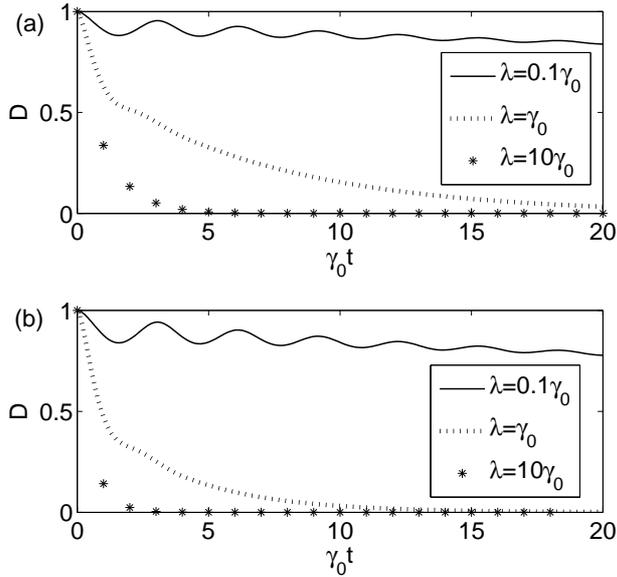}
\caption{Time evolution of quantum discord for different Lorentzian widths $\lambda$ and for initial states: (a) $|\phi\rangle=1/\sqrt{2}[|01\rangle+e^{i\delta}|10\rangle]$,
  and (b) $|\psi\rangle=1/\sqrt{2}[|00\rangle+e^{i\theta}|11\rangle]$. Where $\Delta=2\gamma_{0}$.}
\label{fig:5}
\end{figure}

At the end of this section, we point out that the protection of
quantum correlations may be distinguished as two cases. In the first
case, the quantum correlations in a given state decrease very slowly
in time, so that in a relatively long time they remain on a very
high level, though they may vanish eventually. This protection for
quantum correlations is usually necessary in practice. What we
studied above belong to this case. In the second case, the system
evolves to a steady state in which some residual quantum
correlations exist even for infinitely long time, though the amount
of the quantum correlations may not be very high. We may visually
call the latter case the "correlation trapping". It is worthwhile to
point out that though the phenomenon of "entanglement trapping" in
the local non-Markovian environments has been found
already\cite{Bellomo1,Tong}, the study about preservation of discord
is not trivial, because quantum discord may has different behaviors,
such as the robustness against noises \cite{Werlang,Wang}, and the
distinct features at phase transition
\cite{Dillenschneider,Sarandy}, etc.

\section{Conclusion}
\label{sec:5}
The exact non-Markovian dynamics of quantum-correlations (quantum
entanglement and discord) for two qubits independently coupled to
their respective structured reservoirs (Lorentzian spectrum) have
been investigated. We have found, under some parameter conditions,
that non-Markovian effect can lead to quantum correlation
oscillations which can further interfere to form correlation quantum
beats. This kind of quantum interference may be used to detect the
difference of the qubit-transition frequencies or to indicate the
difference of the local non-Markovian environments. We have also
discussed the preservation of quantum correlations by the effective
suppression of the spontaneous emission. Good monochromaticity of
cavity mode, and large frequency detuning between qubit and cavity
mode can help to protect quantum discord and entanglement.

The intrinsic character of non-Markovian processes with memory lies
in the back flow of information from environment to the system,
which compared to Markovian processes gives rise to various distinct
dynamical traits, such as the changes of dissipative \cite{Breuer3}
or dephasing \cite{Mogilevtsev} rates, the correlation of photons
emitted by an atom \cite{Dubin}, and the non-continuously reduction
\cite{Bellomo} of quantum entanglement or discord, etc. In this
paper, we have further found that the correlation oscillations
induced by local non-Markovian environments can take place
interference to present correlation quantum beats. This discovery
reveals further the dynamical features of quantum correlations and
non-Markovian processes.

The observation of correlation quantum beats requires to enter
strong coupling regime ($\gamma_{0}>\lambda/2$) and under large
detuning conditions ($\Delta>>\lambda$). In recent experiments with
quantum dot micropillars \cite{Reitzenstein} , the quality factor up
to $1.65\times10^{4}$ has been reported and a linewidth $\lambda$ of
$9.6\mu eV$ for the fundamental mode (resonant frequency
$\omega_{c}/2\pi=1.31832eV$ ) has been reached. The linewidth
$\gamma_{0}$ of dot exciton can be changed from about $1.5 \mu eV$
at zero temperature to  $3 meV$ at room temperature \cite{Bayer}.
Thus the strong coupling condition could be well met by temperature
control. The large detuning condition is easy to be realized. Noting
that $50\lambda=480\mu eV$ and the nearest spectator mode with
frequency $1.32070eV$, thus one can make the dot transition
frequency $\omega_{0}$ to be large detuning from the cavity
fundamental mode while simultaneously be immune from the spectator
mode. It also need to note that the entanglement evolution in
non-Markovian environments was studied experimentally, and ESD and
entanglement birth were observed \cite{Xu1}. These achievements have
paved the way to experimentally simulate the paradigmatic models of
open quantum systems.





\end{document}